\documentclass[aps]{revtex4}
\usepackage[dvips]{graphicx}
\usepackage{amsmath}

\begin{document}
\title{Gauging dual symmetry}
\author{Akira Kato}
\email{ak086@csufresno.edu}
\author{Douglas Singleton}
\email{dougs@csufresno.edu}
\affiliation{Physics Dept., CSU Fresno, 2345 East San Ramon Ave.
M/S 37 Fresno, CA 93740-8031, USA}

\date{\today}

\begin{abstract}
The idea of gauging ({\it i.e.} making local) symmetries
is a central feature of many modern
field theories. Usually, one starts with a Lagrangian for
some scalar or spinor fields, with the Lagrangian being
invariant under some global symmetry transformation of these
fields. Making this global symmetry local requires
the introduction of vector fields. Therefore the
vector field is a consequence or result of the gauge
principle. Here we show that for some symmetries the process
of transforming from a global to a local symmetry can be
achieved by introducing a scalar rather than a vector field.
The symmetry that we study is electric-magnetic dual
symmetry which ``rotates'' electric and magnetic quantities into
one another. Starting from an initial Lagrangian which contains
vector fields and satisfies a global electric-magnetic duality,
we show that it is possible to make the symmetry local by
introducing a scalar field.
\end{abstract}


\maketitle

\section{Introduction}
Local or gauge symmetry is deeply ingrained in
modern physics. The strong and electroweak interactions of
particle physics are formulated as gauge interactions. General
relativity can be viewed as taking the global spacetime symmetries
of special relativity and making them local. In the case
of the strong and electroweak interactions the vector fields
can be said to be derived from the gauge principle, in
contrast to the matter fields which are ``put in by hand''.
Here we show that it is possible to gauge certain symmetries
using scalar rather than vector fields. Starting with a
Lagrangian with a global electric-magnetic
dual symmetry and vector fields, we find that
making this dual symmetry local requires the introduction of a
complex scalar field. The final Lagrangian contains both the
original vector fields as well as the scalar field which arises
from the alternative gauge principle. This Lagrangian is different
from the standard scalar electrodynamics Lagrangian in that
the coupling between the vector and scalar field is
a derivative coupling as opposed to a polynomial coupling.

For definiteness and in order to make comparisons, we briefly
review the textbook example \cite{ryder} of scalar electrodynamics
where the ordinary gauge principle is applied to a complex scalar
field. Starting with the Lagrange density
\begin{equation}
\label{1-1}
{\cal L}_{scalar} = (\partial _{\mu} \phi) ^{\ast} (\partial ^{\mu} \phi)
-m^2 \phi ^{\ast} \phi - \lambda (\phi ^{\ast} \phi)^2
\end{equation}
one finds that it is possible to allow the scalar fields to have
the following, local symmetry
\begin{equation}
\label{1-2}
\phi (x) \rightarrow e^{-ie \Lambda (x)} \phi (x)
\; \; \; \;
\phi ^{\ast} (x) \rightarrow e^{ie \Lambda (x)} \phi ^{\ast} (x)
\end{equation}
if one introduces a four-vector, gauge field $A_{\mu}$ which
promotes the ordinary derivative to a covariant derivative --
$\partial _{\mu} \rightarrow \partial _{\mu} -i e A_{\mu} \equiv D_{\mu}$.
In conjunction with the transformation in Eqs. (\ref{1-2})
$A_{\mu}$ transforms as
\begin{equation}
\label{1-3}
A_{\mu} \rightarrow A_{\mu} -\partial _{\mu} \Lambda (x)
\end{equation}
One also introduces a term which contains only the gauge
fields and is invariant under Eq. (\ref{1-3}) namely
$-\frac{1}{4} F_{\mu \nu} F^{\mu \nu}$ where
$F_{\mu \nu} =\partial _{\mu} A_{\nu} - \partial _{\nu} A_{\mu}$.
The new Lagrange density which is invariant under
Eqs.(\ref{1-2}) (\ref{1-3}) is
\begin{equation}
\label{1-4}
{\cal L'}_{scalar} = (D_{\mu} \phi) ^{\ast} (D ^{\mu} \phi)
-m^2 \phi ^{\ast} \phi - \lambda (\phi ^{\ast} \phi)^2
-\frac{1}{4} F_{\mu \nu} F^{\mu \nu}
\end{equation}
Starting with a Lagrange density with scalar
fields it is necessary to introduce a vector field
in order to allow the global phase symmetry of the scalar
fields to become local. 

\section{Dual symmetry}

Source-free electromagnetism possesses a dual symmetry
between electric and magnetic fields which can be
written in terms of $F_{\mu\nu}$ and its dual
${\tilde F}_{\mu \nu} =\frac{1}{2} \epsilon _{\mu \nu \alpha \beta}
F^{\alpha \beta}$ \cite{fels}
\begin{eqnarray}
\label{2-1}
F_{\mu \nu} \rightarrow \cos(\Lambda ) F_{\mu \nu} + \sin(\Lambda )
{\tilde F}_{\mu \nu}
\\
{\tilde F}_{\mu \nu} \rightarrow -\sin(\Lambda ) F_{\mu \nu} + \cos(\Lambda )
{\tilde F}_{\mu \nu} \nonumber
\end{eqnarray}
The standard form of this dual symmetry in terms of the
${\bf E}$ and ${\bf B}$ fields is obtained by making the replacements
$F_{\mu \nu} \rightarrow {\bf E}$ and ${\tilde F}_{\mu \nu} \rightarrow
{\bf B}$ \cite{jackson}. This dual symmetry of Maxwell's equations
can be extended to the case with sources if one allows both electric
and magnetic charges. We would like to extend the dual
symmetry of Eq. (\ref{2-1}) down to the level of the four-vector
potential, $A_{\mu} = (\Phi , {\bf A})$. However, since the
expression for the electric and magnetic fields in terms of the
potentials is not symmetric (${\bf E} = -\nabla \Phi - \partial _t {\bf A}$
and ${\bf B} = \nabla \times {\bf A}$) this is difficult. In the
context of electromagnetism with magnetic charge one can introduce
a second four-vector potential, $C_{\mu} = (\Phi _m , {\bf C})$,
\cite{cabibbo} \cite{zwang} in terms of which the electric and
magnetic fields take on the more symmetric form, 
${\bf E} = -\nabla \Phi - \partial _t {\bf A} -\nabla \times {\bf C}$
and  ${\bf B} = -\nabla \Phi_m - \partial _t {\bf C} +
\nabla \times {\bf A}$. Ref. \cite{sing1} contains an elementary
overview of this two-potential approach, as well as
references to the extensive work that has been done on this
approach to electromagnetism with magnetic charge. In this two-potential
approach the dual symmetry can be extended to the level of
the potentials by changing $F_{\mu \nu} \rightarrow A_{\mu}$ and
${\tilde F}_{\mu \nu} \rightarrow C_{\mu}$ in Eq. (\ref{2-1}) (see
Eq. (5.17b) in Ref. \cite{zwang}).
This global, dual symmetry for the potentials can be put in
a form similar to the phase transformation of the scalar fields
of Eq. (\ref{1-2}) of the previous section by defining a complex
four-potential, $W_{\mu} = A_{\mu} + i C_{\mu}$, in terms of which
Eq. (\ref{2-1}) for the potentials can be written as
\begin{equation}
\label{2-2}
W_{\mu} \rightarrow e^{-i \Lambda} W_{\mu}
\end{equation}
In this form the electric-magnetic dual symmetry is
similar to the phase symmetry of the scalar fields in
Eq. (\ref{1-2}). $W_{\mu}$ gives a complex field
strength tensor, $G_{\mu \nu} = \partial_{\mu} W_{\nu} -
\partial _{\nu} W_{\mu}$, which will be useful in the next section.
Although, we have associated the transformation
in Eq. (\ref{2-2}) with the electric-magnetic duality of Maxwell's
equations, one could argue that Eq. (\ref{2-2}) is just a phase
transformation for a complex, matter, vector field, $W_{\mu}$. However,
in the next section we will give a procedure for making this symmetry
local which is distinct from the standard gauge procedure. Applying
the standard gauge procedure to $W_{\mu}$ would simply lead
to another vector field (this is done, for example, on pg. 124 of
Ref. \cite{fels}.) In contrast our dual gauge procedure will
lead to the introduction of a scalar field. Thus, regardless of
the interpretation of the symmetry in Eq. (\ref{2-2}) the gauging
procedure presented in the next section is different from the
standard method of gauging a symmetry.

\section{Making dual symmetry local}

We now want to allow the dual symmetry of Eq. (\ref{2-2}) to become
local, $\Lambda \rightarrow g \Lambda (x)$. We have introduced the
constant, $g$, which will be seen to be the coupling constant
between the vector field,$W_{\mu}$, and the scalar field.
In our development we build up our Lagrange density one piece
at a time using an infinitesimal version ({\it i.e.} taking
$\Lambda (x)$ infinitesimal) of
Eq. (\ref{2-2}) namely
\begin{equation}
\label{3-1}
\delta W_{\mu} =-i g \Lambda W_{\mu}
\; \; \; \;
\delta W_{\mu} ^{\ast} = i g \Lambda W_{\mu } ^{\ast}
\end{equation}
We will also need the infinitesimal variations of the
partial derivatives of the complex vector potential
\begin{equation}
\label{3-2}
\delta (\partial _{\mu} W_{\nu}) =-i g \partial _{\mu} (\Lambda W_{\nu})
\; \; \; \;
\delta (\partial _{\mu} W_{\nu} ^{\ast}) =
i g \partial_{\mu} (\Lambda  W_{\nu } ^{\ast} )
\end{equation}
and the variations of the complex field strengths
\begin{eqnarray}
\label{3-3}
\delta G_{\mu \nu} &=& 
-i g \Lambda G_{\mu \nu} -ig (\partial _{\mu} \Lambda W_{\nu}
-\partial _{\nu} \Lambda W_{\mu})
\nonumber \\
\delta G_{\mu \nu} ^{\ast} &=& 
i g \Lambda G_{\mu \nu} ^{\ast} + ig (\partial _{\mu} \Lambda
W_{\nu} ^{\ast} -\partial _{\nu} \Lambda W_{\mu}^{\ast})
\end{eqnarray}
Now we start with a ``kinetic'' energy Lagrangian for the
vector fields
\begin{equation}
\label{3-4}
{\cal L} _1 = -\frac{1}{4} G_{\mu \nu} G^{\mu \nu \ast}
\end{equation}
in the same way that in the introduction we began with
a kinetic energy term for the complex scalar field.
This Lagrangian in Eq. (\ref{3-4}) is invariant under a generalized
version of the gauge transformation in Eq. (\ref{1-3}) ({\it i.e.}
a transformation for both $A_{\mu}$ and $C_{\mu}$). The final
Lagrange density that we find, will no longer respect this
standard gauge symmetry, but it will be invariant under the
local version of the dual symmetry in Eq. (\ref{2-2}). Thus
we gain one local symmetry, Eq. (\ref{2-2}), at the cost of
losing another, Eq. (\ref{1-3}). For
the scalar field case we also included a mass and quartic
self interaction term since these were allowed by the phase
symmetry of Eq. (\ref{1-2}). In the same way we could include
a mass term, $m^2 W_{\mu} W^{\mu \ast}$, and self interaction term,
$\lambda (W_{\mu} W^{\mu \ast})^2$ to our Lagrangian. Such terms
are usually forbidden by the standard gauge transformation in
Eq. (\ref{1-3}), but are allowed by Eq. (\ref{2-2}). We could
also add a term like $\epsilon _{\mu \nu \alpha \beta} G^{\mu \nu}
G^{\alpha \beta \ast} = G^{\mu \nu} {\widetilde G}_{\mu \nu}^{\ast}$ to
${\cal L}_1$. However, by the anti-symmetry properties
of $\epsilon_{\mu \nu \alpha \beta}$,
such a term would not change the field equations
derived from the Lagrangian. In the end when we arrive
at the dual version of the covariant derivative it will be
straightforward to show that both terms like $G_{\mu \nu}
G^{\mu \nu \ast}$ and $G^{\mu \nu} {\widetilde G}_{\mu \nu}^{\ast}$ can
be made consistent with the local dual symmetry of
Eq. (\ref{2-1}). Taking the variation of ${\cal L}_1$ using
Eq. (\ref{3-3})
\begin{eqnarray}
\label{3-5}
\delta {\cal L}_1 &=& -\frac{1}{4}(\delta G_{\mu \nu} G^{\mu \nu \ast}
+G_{\mu \nu} \delta G^{\mu \nu \ast}) \nonumber \\
&=& \frac{i g}{2} \partial ^{\mu} \Lambda
(W^{\nu} G_{\mu \nu} ^{\ast} - W^{\nu \ast} G_{\mu \nu})
\end{eqnarray}
Since $\delta {\cal L} _1 \ne 0$ we continue to
add terms to the Lagrangian in the hopes of building
a total Lagrangian for which $\delta {\cal L} _{total} = 0$.
We next consider
\begin{equation}
\label{3-6}
{\cal L} _2 = \frac{g}{2}(\partial _{\mu} \phi W_{\nu} G^{\mu \nu \ast}
+\partial _{\mu} \phi ^{\ast} W_{\nu} ^{\ast} G^{\mu \nu})
\end{equation}
where we have introduced a complex, scalar field $\phi$ which
we require to transform as
\begin{equation}
\label{3-7}
\phi \rightarrow \phi - i \Lambda (x)
\; \; \; \;
\phi^{\ast} \rightarrow \phi ^{\ast} + i \Lambda (x)
\end{equation}
The arbitrary function, $\Lambda (x)$, is the same
as in Eq. (\ref{2-2}). Just as the dual transformation of
Eq. (\ref{2-2}) was similar to the phase transformation of
Eq. (\ref{1-2}), so the transformation of Eq. (\ref{3-7})
is similar to the gauge transformation of Eq. (\ref{1-3}).
We will call $\phi$ the ``gauge'' field for the dual symmetry or
the dual gauge field. Since the transformation of the scalar
field only involves an imaginary shift of the field via
$i \Lambda (x)$ one could use this freedom to transform
away the imaginary part of the scalar field by choosing
$\Lambda (x)$ to equal the imaginary part of $\phi$.
This freedom will manifest itself later in that the
scalar field kinetic energy term, allowed by Eq. (\ref{3-7}),
will only contain the real part of the scalar field.

The infinitesimal forms of the transformation
for $\phi$ and its partial derivatives are given by
\begin{eqnarray}
\label{3-8}
\delta \phi = - i \Lambda \; \; \; \;
\delta (\partial _{\mu} \phi) = - i \partial _{\mu} \Lambda
\\
\delta \phi ^{\ast} =  i \Lambda \; \; \; \;
\delta (\partial _{\mu} \phi ^{\ast}) =  i \partial _{\mu} \Lambda
\nonumber
\end{eqnarray}
Using these and the transformations of Eqs.
(\ref{3-1}) - (\ref{3-3}) we find
\begin{eqnarray}
\label{3-9}
\delta {\cal L} _2 &=& -\frac{i g}{2} \partial _{\mu} \Lambda
(W_{\nu} G^{\mu \nu \ast} -W_{\nu} ^{\ast} G^{\mu \nu})
+ \frac{i g^2}{2} \partial _{\mu} \Lambda (\partial ^{\mu} \phi
- \partial ^{\mu} \phi ^{\ast}) W_{\nu} ^{\ast} W^{\nu}
\nonumber \\
&+&\frac{i g^2}{2} \partial _{\nu} \Lambda (\partial _{\mu} \phi ^{\ast}
W^{\nu \ast} W^{\mu} -\partial _{\mu} \phi W^{\nu} W^{\mu \ast}) 
\end{eqnarray}
The first term in $\delta {\cal L} _2$ now cancels the
unwanted term from $\delta {\cal L} _1$, but only at the
cost of two new terms. Next we add
\begin{equation}
\label{3-10}
{\cal L}_3 = -\frac{g^2}{2} \partial _{\mu} \phi \partial ^{\mu}
\phi ^{\ast} W_{\nu} W^{\nu \ast}
\end{equation}
which has the following variation from Eqs.
(\ref{3-1}) and (\ref{3-8})
\begin{equation}
\label{3-11}
\delta {\cal L} _3 = -\frac{i g^2}{2} \partial _{\mu} \Lambda 
(\partial ^{\mu} \phi - \partial ^{\mu} \phi ^{\ast})
W_{\nu} ^{\ast} W^{\nu} 
\end{equation}
which cancels the second term from $\delta {\cal L} _2$
in Eq. (\ref{3-9}). In arriving at Eq. (\ref{3-11}) we used
the result that $W_{\nu} ^{\ast} W^{\mu}$ is invariant under
the local dual transformation so that $\delta (W_{\nu} ^{\ast}
W^{\mu}) = 0$. Finally, adding
\begin{equation}
\label{3-12}
{\cal L}_4 = \frac{g^2}{4} (\partial ^{\nu} \phi \partial _{\mu}
\phi ^{\ast} W_{\nu} ^{\ast} W^{\mu} +
\partial ^{\nu} \phi ^{\ast} \partial _{\mu}
\phi W_{\nu} W^{\mu \ast})
\end{equation}
gives a variation of
\begin{equation}
\label{3-13}
\delta {\cal L} _4 =\frac{i g^2}{2} \partial _{\nu} \Lambda
(\partial _{\mu} \phi W^{\nu} W^{\mu \ast} -
\partial _{\mu} \phi ^{\ast} W^{\nu \ast} W^{\mu} )
\end{equation}
where we have renamed indices to get this form. Again, we
have used $\delta (W_{\nu}^{\ast} W^{\mu}) = 0$ and
$\delta (W_{\nu} W^{\mu \ast}) = 0$. This variation
of ${\cal L}_4$ cancels the third term from $\delta {\cal L} _2$
in Eq. (\ref{3-9}), and by adding all four terms together
(${\cal L} = {\cal L}_1 +{\cal L}_2+{\cal L}_3+{\cal L}_4$) we
arrive at a Lagrange density which is invariant under the local
dual transformation ({\it i.e.} $\delta {\cal L} = 0$). In the
standard application of the gauge principle sketched in the
introduction, one finishes by adding a term to the
Lagrange density which contains only the vector fields
({\it i.e.} the last term in Eq. (\ref{1-4})).
The same thing is possible for the local dual symmetry
with the Lagrange density of the form
\begin{equation}
\label{3-14}
{\cal L}_5 =\frac{1}{2}(\partial _{\mu} \phi + \partial _{\mu} \phi ^{\ast})
(\partial ^{\mu} \phi + \partial ^{\mu} \phi ^{\ast})
\end{equation}
It is easy to see that under the infinitesimal transformation,
Eq. (\ref{3-8}), $\delta {\cal L}_5 =0$. The Lagrange density
of Eq. (\ref{3-14}) looks similar to the standard
kinetic energy terms of a scalar field namely
$\frac{1}{2} \partial _{\mu} \phi \partial ^{\mu} \phi$. Collecting
these five terms together the total Lagrange density can be written
in the simplified form
\begin{eqnarray}
\label{3-15}
{\cal L}_{total} &=& -\frac{1}{4}(G_{\mu \nu} - g \partial _{\mu} \phi
W_{\nu}+ g \partial _{\nu} \phi W_{\mu})
(G^{\mu \nu \ast} - g \partial ^{\mu} \phi ^{\ast} W^{\nu \ast}
+ g \partial ^{\nu} \phi ^{\ast} W^{\mu \ast}) \\
&+&\frac{1}{2}(\partial _{\mu} \phi + \partial _{\mu} \phi ^{\ast})
(\partial ^{\mu} \phi + \partial ^{\mu} \phi ^{\ast})  \nonumber
\end{eqnarray}
This is invariant ($\delta {\cal L}_{total} = 0$) under the dual
``gauge'' transformations of Eqs. (\ref{3-1}) (\ref{3-2}) (\ref{3-8}).
The scalar-vector theory associated with the Lagrange density of
Eq. (\ref{3-15}) is distinct from scalar electrodynamics. The scalar-vector
couplings given in Eqs. (\ref{3-6}) (\ref{3-10}) (\ref{3-12})
are all derivative couplings, whereas scalar electrodynamics also has
polynomial couplings between the scalar and vector field. The
coupling $g$ in the Lagrangian of Eq. (\ref{3-15}) also has
a mass dimension $-1$, whereas
scalar electrodynamics has a dimensionless
coupling. If the Lagrange density in Eq. (\ref{3-15}) is to have
a mass dimension $4$, and if $\phi$, $W_{\mu}$ and $\partial_{\mu}$
have the conventional mass dimension of $1$, then $g$ must have
mass dimension $-1$. This last comment raises the question
as to the renormalizability of the scalar-vector theory of
Eq. (\ref{3-15}). The fact that $g$ has a negative mass dimension
indicates that the theory associated with Eq. (\ref{3-15}) is
non-renormalizable. However, theories which are non-renormalizable
can still be useful when treated as effective theories
\cite{wein}. In any case for the present paper we are focused
on the task of constructing a classical Lagrangian which
respects the dual symmetry locally. We leave the technical
and complex question of the renormalization of the theory
in Eq. (\ref{3-15}) for a possible future investigation.

The derivative coupling which arises between the vector and scalar
fields from the dual gauge principle (Eqs. (\ref{3-6})
(\ref{3-10}) (\ref{3-12})) can be compared to the derivative
couplings which occur in an effective Lagrangian for pions \cite{don}
\begin{equation}
\label{3-16}
{\cal L}_{eff} = \frac{1}{2} \partial _{\mu} {\vec \pi}
\partial^{\mu} {\vec \pi} + \frac{1}{6F^2}\left[ ({\vec \pi}\cdot
\partial _{\mu} {\vec \pi})^2 - {\vec \pi}^2 (\partial _{\mu}
{\vec \pi} \cdot \partial ^{\mu} {\vec \pi}) \right] +.....
\end{equation}
where $F$ is a coupling constant with mass dimension 1, and
${\vec \pi}$ is the pion triplet. Here the derivative couplings are
between scalar fields, while in Eq. (\ref{3-15}) the couplings are between
scalar and vector fields. Also just as with the Lagrangian
in Eq. (\ref{3-15}) the effective Lagrangian in Eq. (\ref{3-16}) is
non-renormalizable.

Although, simple mass terms, like $m^2 \phi \phi ^{\ast}$, are
forbidden by the dual gauge
transformation (\ref{3-7}), one can use the invariance
of $\phi + \phi ^{\ast}$ to add a term like
$m^2 (\phi + \phi ^{\ast})^2$, which is a mass term for
the real part of the scalar field. Writing out the complex scalar dual
gauge field in terms of real components ($\phi = \varphi _1 + i \varphi _2$)
one notices the  combination, $(\phi + \phi ^{\ast})$, contains
only one real degree of freedom, $\varphi _1$. This comment also
applies to $\partial _{\mu} \phi + \partial _{\mu}
\phi ^{\ast}$ which appears in the ``kinetic'' energy term
for the scalar field, Eq. (\ref{3-14}). Therefore, even though it
appears that there are two scalar degrees of freedom associated with
the complex scalar field, $\phi$, only one degree of freedom,
$\varphi _1$, has a proper kinetic energy or mass term. This
result that only one degree of freedom ({\it i.e.} the real
component) from the complex scalar field appears to be dynamical
is related to the form of transformation of the scalar field
given in Eq. (\ref{3-7}). Making use of this ``gauge'' freedom
one can chose $\Lambda (x) = \varphi _2 (x)$ thus transforming away
the complex degree of freedom of $\phi$.
There are other terms which could be added to ${\cal L}_{total}$ that
preserve the dual gauge invariance: $(\phi +
\phi ^{\ast}) ^n$, $(W_{\mu} W^{\mu \ast})^m$, or
$(\phi + \phi ^{\ast}) ^n (W_{\mu} W^{\mu \ast})^m$, where $n$ and $m$
are arbitrary integers.

We have laboriously arrived at this Lagrange density
using the infinitesimal form of the dual gauge transformation.
In the next section we will show how the first term in
${\cal L} _{total}$ can be interpreted as a dual covariant
derivative, leading much more quickly
and elegantly to the invariance of the Lagrange density under the
local dual transformations.

\section{Comparison with ordinary gauge symmetry}

In the previous section we have shown that there is a close
connection between the standard gauge principle and the
dual gauge principle that arises by making the electric-magnetic
dual symmetry local. The difference is that the scalar and vector
fields have switched roles, and are thus in some sense duals
of one another. In an ordinary
gauge theory one has the phase transformation
$\phi \rightarrow e^{-i e \Lambda (x)} \phi$,
and the gauge transformation $A_{\mu} \rightarrow A_{\mu}
- \partial _{\mu} \Lambda$. For the dual gauge theory one
has the local, dual symmetry $W_{\mu} \rightarrow e^{-i g \Lambda (x)}
W_{\mu}$ and the dual gauge transformation $\phi \rightarrow \phi
-i \Lambda (x)$. In an ordinary gauge theory one starts
with scalar or spinor fields and introduces vector fields in
order to have the phase symmetry of the scalar or
spinor fields become local. For the dual gauge
symmetry one starts with vector fields and introduces 
scalar fields so that the dual symmetry of the vector fields
can become local. In an ordinary gauge theory one introduces
a kinetic term for the vector fields, $-\frac{1}{4} F_{\mu\nu}
F^{\mu \nu}$, which is invariant by itself under the gauge transformation
of $A_{\mu}$. In the dual gauge theory one introduces a kinetic
term for the dual gauge fields, $(\partial _{\mu} \phi
+ \partial _{\mu} \phi ^{\ast})( \partial ^{\mu} \phi
+\partial ^{\mu} \phi ^{\ast})$, which is invariant by itself
under the dual gauge transformation of $\phi$.
In an ordinary gauge theory one starts with
a kinetic energy Lagrangian for the matter fields
$\partial _{\mu} \phi \partial ^{\mu} \phi ^{\ast}$, and the
interaction terms between the matter and gauge fields arise as
a result of the replacement of the ordinary derivative by the
covariant derivative, $\partial _{\mu} \phi \rightarrow
\partial _{\mu} \phi - ie A_{\mu} \phi$. In the dual gauge theory
one can introduce a similar concept. Changing the derivatives
of the vector fields in the following way
\begin{eqnarray}
\label{4-1}
\partial _{\mu} W_{\nu} &\rightarrow& \partial _{\mu} W_{\nu}
-g \partial _{\mu} \phi W_{\nu}
\nonumber \\
\partial _{\mu} W_{\nu}^{\ast} &\rightarrow& \partial _{\mu}
W_{\nu} ^{\ast} - g \partial _{\mu} \phi ^{\ast} W_{\nu} ^{\ast}
\end{eqnarray}
one has
\begin{eqnarray}
\label{4-2}
G_{\mu \nu} &\rightarrow& G_{\mu \nu} - g \partial _{\mu} \phi W_{\nu}
+ g \partial _{\nu} \phi W_{\mu}
\nonumber \\ 
G_{\mu \nu}^{\ast} &\rightarrow& G_{\mu \nu} ^{\ast}
- g \partial _{\mu} \phi^{\ast} W_{\nu} ^{\ast}
+ g \partial _{\nu} \phi ^{\ast} W_{\mu}^{\ast}
\end{eqnarray}
which can be seen to give the first term in the Lagrangian
of Eq. (\ref{3-15}) starting from $-\frac{1}{4} G_{\mu \nu}
G^{\mu \nu \ast}$. In addition one can show that
the dual ``covariant'' derivative in Eq. (\ref{4-1}) transforms
as
\begin{eqnarray}
\label{4-3}
\partial _{\mu} W_{\nu} - g \partial _{\mu} \phi
W_{\nu}  \rightarrow
e^{-i g \Lambda(x)} (\partial _{\mu} W_{\nu} - g \partial _{\mu}
\phi W_{\nu})
\\
\partial _{\mu} W_{\nu} ^{\ast} - g \partial _{\mu} \phi ^{\ast}
W_{\nu} ^{\ast} \rightarrow
e^{i g \Lambda(x)} (\partial _{\mu} W_{\nu} ^{\ast} - g \partial _{\mu}
\phi ^{\ast} W_{\nu} ^{\ast})
\end{eqnarray}
The dual covariant derivative also
provides an easier and more elegant way of seeing that terms like
$G_{\mu \nu} G^{\mu \nu \ast}$ (and also terms like $G_{\mu \nu}
\widetilde{G}^{\mu \nu \ast}$) can be made consistent with the local
dual symmetry. 

Finally in ordinary gauge theories mass terms
for the vector gauge bosons, $m^2 A_{\mu} A^{\mu}$, are forbidden
by the gauge transformation.
Under the dual gauge transformation mass terms and
self interaction terms are allowed for the vector fields.
One can have terms like $W_{\mu} W^{\mu \ast}$ or
$(W_{\mu} W^{\mu \ast})^2$ which are invariant under the
dual gauge transformation. Similar quartic terms for the
vector fields can arise in ordinary gauge theories if the
gauge symmetry is non-Abelian. In our example of the preceding
section we started with an Abelian theory in Eq. (\ref{3-4}), which
satisfied an ordinary gauge symmetry -- Eq. (\ref{1-3}). Our final
Lagrangian -- Eq. (\ref{3-15}) -- no longer satisfied this original
gauge symmetry, but instead satisfied a local, dual symmetry. Thus
one local symmetry has been lost or exchanged in favor of another.

\section{Conclusions}

By gauging the electric-magnetic dual
symmetry of Maxwell's field equations the roles of the
vector fields and the scalar fields are to some extent
exchanged. In the dual gauge theory the scalar fields arise
from the dual gauge principle, in the same way that in ordinary
gauge theories the vector fields
arise from the ordinary gauge principle.
There are other interesting formulations of the ordinary
gauge principle \cite{chaves} where scalar gauge fields arise
in {\it conjunction} with the usual vector gauge fields. For the dual
gauge theory, however, the roles of the scalar and vector fields
are exchanged. There have been other recent attempts
\cite{pak} to make the dual symmetry of the Schwarz-Sen electromagnetic
action \cite{sen} local. In Ref. \cite{pak} this is done
without the introduction of a scalar field. There are two
obvious extensions of the dual gauge idea:

\begin{enumerate}

\item{Non-Abelian theories have been shown to have dualities similar
to the electric-magnetic duality of the Abelian Maxwell equations
\cite{tsou}. Thus, one could consider gauging the dual symmetry for
a non-Abelian gauge theory. However, there are significant differences
between the Abelian electric-magnetic duality discussed here,
and the non-Abelian version given in Ref. \cite{tsou}.}

\item{In our present example of the dual gauge idea we have been able
to ``derive'' scalar fields from the gauging of the dual symmetry.
One could ask if it is possible to ``derive'' fermionic fields
from the dual gauge idea. In our example the scalar fields
were first introduced in Eq. (\ref{3-6}) via ${\cal L}_2$. The
dual gauge transformations of these scalar fields were then chosen
as that the first term in $\delta {\cal L}_2$, which arose from the
variation of the scalar fields, would cancel $\delta {\cal L}_1$.
To accomplish the same thing with fermionic field we would need to
replace $\partial _{\mu} \phi$, $\partial _{\mu} \phi ^{\ast}$ in Eq.
(\ref{3-6}) with fermionic terms which also have one Lorentz index
({\it e.g.} $\partial _{\mu} \phi$, $\partial _{\mu} \phi ^{\ast}$
$\rightarrow$ $\partial_{\mu} [{\bar \psi} \psi$] or
${\bar \psi} \gamma _{\mu} \psi$). The fermionic fields would then
need to satisfy some transformation akin to Eq. (\ref{3-7}) so that
the variation of the fermionic fields would cancel
$\delta {\cal L}_1$.}

\end{enumerate}

The dual gauge principle given in this article replaces
the vector, gauge field with the derivative of a scalar field in
the definition of the covariant derivative ({\it i.e.} $\partial_{\mu}
-ie A_{\mu} \rightarrow \partial _{\mu} -g \partial_{\mu} \phi$ or
$i e A_{\mu} \rightarrow g \partial _{\mu} \phi$. There are other cases
when a vector, gauge field can be identified with the derivative
of scalar fields. For pure SU(2) gauge theory it is possible to
make the following ansatz \cite{corr} \cite{wilc}
\begin{equation}
\label{5-1}
A_{\mu}^a=(\epsilon_{0 a \mu \nu} \mp i g_{a \mu}
g_{\nu 0}\pm i g_{a \nu} g_{\mu 0}) \frac{\partial ^{\nu}
\phi}{\phi}  =\eta _{a\mu \nu} \frac{\partial ^{\nu} \phi}{\phi}
\end{equation}
where $\epsilon_{\alpha \beta \mu \nu}$ is the 4D Levi-Civita
symbol, and $g_{\mu \nu}$ is the metric tensor. 
This ansatz turns the SU(2) Yang-Mills theory into a
massless $\phi ^4$ theory. The
relationship between the vector gauge field and scalar
field given in Eq. (\ref{5-1}) is more complicated than
the relationship implied by the comparison of the ordinary
and dual gauge covariant derivative ({\it i.e.}
$i e A_{\mu} \rightarrow g \partial _{\mu} \phi$).
Nevertheless both relationships involve the association/replacement
of a vector field by the partial derivative of a scalar
field.

\section{Acknowledgment} The authors are grateful to Dr. Gerardo
Mu{\~n}oz and Dr. Ruben Pakman for discussions and comments on
this work.

\end{document}